\documentclass[aps,prb,twocolumn,floatfix]{revtex4}
\usepackage{amsmath,amssymb,epsfig}

\begin{document}


\title{The screening of $4f$ moments and delocalization in the compressed light
rare earths}

\author{A. K. McMahan,$^1$ R. T. Scalettar,$^2$ and M. Jarrell$^3$}
\affiliation{$^1$Physical and Life Sciences Directorate, Lawrence
Livermore National Laboratory, Livermore, CA 94550\\ $^2$Physics
Department, University of California, Davis, CA 95616\\ $^3$Department of
Physics and Astronomy, Louisiana State University, Baton Rouge, LA 70803}

\date{\today}

\begin{abstract} 
Spin and charge susceptibilities and the $4f^n$, $4f^{n-1}$, and
$4f^{n+1}$ configuration weights are calculated for compressed Ce
($n\!=\!1$), Pr ($n\!=\!2$), and Nd ($n\!=\!3$) metals using dynamical
mean field theory combined with the local-density approximation.  At
ambient and larger volumes these trivalent rare earths are pinned at sharp
$4f^n$ configurations, their $4f$ moments assume atomic-limiting values,
are unscreened, and the $4f$ charge fluctuations are small indicating
little $f$ state density near the Fermi level.  Under compresssion
there is dramatic screening of the moments and an associated increase
in both the $4f$ charge fluctuations and static charge susceptibility.
These changes are coincident with growing weights of the $4f^{n\!-\!1}$
configurations, which it is argued are better measures of delocalization
than the $4f^{n\!+\!1}$ weights which are compromised by an increase
in the number of $4f$ electrons caused by rising $6s$, $6p$ bands.
This process is continuous and prolonged as a function of volume,
with strikingly similarity among the three rare earths, aside from
the effects moderating and shifting to smaller volumes for the heavier
members.  The observed $\alpha$-$\gamma$ collapse in Ce occurs over the
large-volume half of this evolution, the Pr analog at smaller volumes,
and Nd has no collapse.
\end{abstract}



\maketitle

\section{Introduction}


The trivalent rare earth series is an important but not well understood
test ground for the study of strong electron correlation, and the
manner in which its effects diminish as the $4f$ electrons delocalize
under pressure.  These metals appear initially to remain localized as
they pass through a sequence of high-symmetry close packed phases keyed
to $3d$-band occupancy,\cite{Duthie77} and then on further compression,
eventually reach low-symmetry early-actinide-like structures suggestive of
$f$-electron bonding.\cite{Benedict93,Holzapfel95,JCAMD} Transitions in
the region between these two limits may exhibit unusually large volume
changes.  While not the case for Nd,\cite{AkellaNd99,ChesnutNd00}
the $\gamma$-$\alpha$ ``volume collapse'' of 15\% in Ce is well
known,\cite{KoskimakiCe78,OlsenCe85,McMahonCe97,VohraCe99}
and similar collapses occur in Pr
(9\%),\cite{MaoPr81,ZhaoPr95,ChesnutPr00,BaerPr03,CunninghamPr05}
Gd (5\%),\cite{HuaGd98} and Dy (6\%).\cite{PattersonDy04} Magnetic
properties at atmospheric pressure are generally consistent with
atomic $4f$ Hund's rules moments,\cite{mcewen78,JCAMD} while the
susceptibility for the collapsed $\alpha$-Ce phase\cite{KoskimakiCe78}
and for early actinide analogs\cite{ward86} is temperature-independent,
enhanced Pauli paramagnetic, indicating absent or screened moments.
On the other hand, high energy neutron scattering measurements for
Ce,\cite{Murani05} and x-ray emission spectroscopy for Gd,\cite{Maddox06}
continue to detect $4f$ moments in the collapsed phases, possibly
sensing underlying ``bare'' moments in spite of screening effects.
The $4f$ electron delocalization itself may be examined using resonant
inelastic x-ray scattering determination of the probabilities of finding
$f^{n\pm1}$ configurations in a compressed rare earth of nominal $f^n$
character.\cite{Maddox06,Rueff06}

There are at present two viable explanations for the Ce collapse,
with possible implications for the other trivalent rare earths. One
is that it is driven by a Mott Transition (MT) in the $4f$
electrons,\cite{Johansson74,Johansson75} while the other Kondo
Volume Collapse (KVC) model points to rapid volume-dependent
changes in screening of the $4f$ moments by the valence
electrons.\cite{KVC1,KVC2,KVC3,KVC4} The conflict between these
scenarios is exaggerated by the use of incompatible approximations.
Polarized local-density approximation (LDA), LDA+U, and self-interaction
corrected LDA calculations have been used to support the MT
picture.\cite{Eriksson90,Sandalov95,Shick01,SoderlindPr02,SzotekCe94,SvaneCe94,SvanePr97}
While valuable, these are still {\it static} mean-field treatments
which yield either completely itinerant (no $4f$ Hubbard structure)
or completely localized (Hubbard splitting but no Fermi-level $4f$
structure) solutions, thus indicating a too abrupt picture of the
collapse transitions.  On the other hand, the Anderson impurity model
treatments\cite{KVC1,KVC2,KVC3,KVC4} used to elucidate the KVC scenario
can be faulted for omitting direct $f$-$f$ hybridization and Kondo
lattice effects, and one may worry whether $O(1/N)$ solutions\cite{1N}
might favor the localized limit.

The combination (LDA+DMFT)\cite{LDADMFT1,LDADMFT2} of LDA
input with truly correlated Dynamical Mean Field Theory
(DMFT)\cite{DMFT1,DMFT2} solutions has offered a new perspective
which has generally been supportive of the KVC scenario for
Ce.\cite{Zoelfl01,Held01,McMahan03,McMahan05,Haule05,Amadon06}
Such calculations for Ce also point to ongoing $4f$ delocalization
in the relevant volume range, a critical driver of Mott
transitions.\cite{Held01,McMahan03}
To further clarify the behavior of
the compressed trivalent rare earths, the present paper reports LDA+DMFT calculations of
the $4f$ spin and charge susceptibilities and the $4f^n$, $4f^{n\pm1}$
configuration weights for the first three members, Ce ($n\!=\!1$),
Pr ($n\!=\!2$), and Nd ($n\!=\!3$).  This work follows an earlier
effort which examined the equation of state and spectra for the same
materials.\cite{McMahan05} Here we confirm that Ce, Pr, and Nd remain
localized at pressures up through the face centered cubic (fcc, $\gamma$
for Ce) phases as indicated by sharp $4f^n$ populations, unscreened
moments with atomic-limiting values, and small charge fluctuations
indicating little $4f$ state density overlapping the Fermi level.
On subsequent compression there is rapid and dramatic screening of the
moments and concurrent increase in charge fluctuations and the static
charge susceptibility.  These changes are also coincident with rapid
growth in the $4f^{n\!-\!1}$ configuration weights, which we argue offer a
truer measure of delocalization than do the $4f^{n\!+\!1}$ weights which
are complicated by the overall increase in the number of $4f$ electrons
due to rising $6s$, $6p$ bands.  These trends are continuous and prolonged
as a function of compression, and strikingly similar among the three rare
earths, suggesting a robust underlying progression which must first be
acknowledged before tackling in general the location or absence of volume
collapse transitions at various stages along the course of this evolution.

In the remainder of this paper, the susceptibility and configuration
weight formalisms are first reviewed in Secs.~II and III, respectively.
Computational details are given in Sec,~IV, results in Sec.~V, and a
summary in Sec.~VI.  The Appendix discusses the optimal disposition of
diagonal, one-body $f$-$f$ terms used here in the Quantum Monte Carlo
solution of auxiliary impurity problem.


\section{Susceptibility}

The important local susceptibilities for real, multiband
systems would appear to be associated with the total spin
{\bf S}, orbital angular momentum {\bf L}, total angular
momentum ${\bf J}={\bf L}+{\bf S}$, and charge
\begin{eqnarray}
\chi_S(\tau)&=&\langle T_\tau\, {\bf \hat{S}}(\tau)\!\cdot\! 
{\bf \hat{S}}(0)\rangle 
\label{chiS} \\
\chi_L(\tau)&=&\langle T_\tau\, {\bf \hat{L}}(\tau)\!\cdot\! 
{\bf \hat{L}}(0)\rangle
\label{chiL} \\
\chi_J(\tau)&=&\langle T_\tau\, {\bf \hat{J}}(\tau)\!\cdot\! 
{\bf \hat{J}}(0)\rangle
\label{chiJ} \\
\chi_c(\tau) &=& \langle T_\tau \, [\hat{n}_l(\tau)\!-\!n_l]
[\hat{n}_l(0)\!-\!n_l] \rangle \, ,
\label{eqnchic}
\end{eqnarray}
for the electrons in some $l$ shell (e.g., $4f$) on a particular site,
and with $T_\tau$ the imaginary time $\tau$ ordering operator.  In the
last, $\hat{n}_l$ is the total number operator $\sum_{m,\sigma}
\hat{n}_{m\sigma}$ with $m=-l,-l\!+\!1,\cdots ,l$, and $n_l$ is its
generally non-integral expectation.  For cubic symmetry,
\begin{eqnarray}
\chi_S(\tau)&=& 3 \, \langle T_\tau\, \hat{S}_z(\tau)\ \hat{S}_z(0)\rangle  
\nonumber \\
&=& \frac34 \sum_{m,\sigma,m^\prime,\sigma^\prime} \sigma \sigma^\prime
\langle T_\tau \, \hat{n}_{m\sigma}(\tau) \,
\hat{n}_{m^\prime\sigma^\prime} (0) \rangle
\nonumber \\
&=& \frac34 
\langle T_\tau [\hat{n}_\uparrow(\tau)\!-\! \hat{n}_\downarrow(\tau)]
[\hat{n}_\uparrow(0)\!-\!\hat{n}_\downarrow(0)]\rangle
\end{eqnarray}
where $\hat{n}_\sigma \equiv \sum_m \hat{n}_{m\sigma}$ is the
total number operator for a given spin $\sigma\equiv 2\,m_s=\pm 1$
summed over orbitals $m\equiv m_l$.  Aside from the factor of $3/4$
this is the spin susceptibility of the two-band model of Koga {\it
et al.},\cite{Koga05} or for $\tau\!=\!0$, the same factor times
the bare local moment $m_z^2$ of one-band Hubbard and Anderson
models. Similarly $\chi_L(\tau)=3\langle T_\tau \hat{L}_z(\tau) 
\hat{L}_z(0) \rangle$ for cubic symmetry and thus
\begin{equation}
\chi_L(\tau)= 
3 \sum_{m,\sigma,m^\prime,\sigma^\prime} m\,m^\prime\,
\langle T_\tau \, \hat{n}_{m\sigma}(\tau) \,
\hat{n}_{m^\prime\sigma^\prime} (0) \rangle \,
\end{equation}
which is $3/4$ times the orbital susceptibility of the
two-band model of Koga {\it et al.},\cite{Koga05} taking
$l\!=\!1/2$ and $m=-1/2,\,1/2$.

\subsection*{Spin susceptibility $\chi_J(\tau)$}

In the presence of the spin-orbit interaction, it may be
more useful to work in a relativistic basis $j=l\pm 1/2$
(except just $j\!=\!1/2$ for $l\!=\!0$), with magnetic
quantum numbers $\nu = -j,\, -j\!+\!1,\cdots,\,j$.  Again for
cubic symmetry,
\begin{equation}
\chi_J(\tau)= 
3 \sum_{j,\nu,j^\prime,\nu^\prime} \nu \,\nu^\prime
\langle T_\tau \, \hat{n}_{j\nu}(\tau) \,
\hat{n}_{j^\prime\nu^\prime} (0) \rangle \, .
\end{equation}
For DMFT calculations which include spin orbit and the
Hubbard repulsion $U$, but omit the Hund's rule intraatomic
exchange terms, a reasonable approximation to the self
energy is $\Sigma_{j,\nu,j^\prime,\nu^\prime}(i\omega)
\sim \delta_{j,j^\prime} \delta_{\nu,\nu^\prime}
\Sigma_j(i\omega)$.  A consistent approximation to $\langle
T_\tau \, \hat{n}_{j\nu}(\tau) \, \hat{n}_{j^\prime\nu^\prime}
(0) \rangle$ is
\begin{eqnarray}
\langle T_\tau \, \hat{n}_{j\nu}(\tau) \, \hat{n}_{j^\prime\nu^\prime} (0)
\rangle \mbox{\hspace{2in}}
\nonumber \\
\sim \left\{ \begin{array}{ll} 
N_j(\tau)/(2j\!+\!1) &  \mbox{if $j\!=\!j^\prime,\,\nu\!=\!\nu^\prime$} \\
D_{jj}(\tau)/[j(2j\!+\!1)]
& \mbox{if $j\!=\!j^\prime,\,\nu\!\neq\!\nu^\prime$} \\
D_{jj^\prime}(\tau)/[(2j\!+\!1)(2j^\prime\!+\!1)]
& \mbox{if $j\!\neq\!j^\prime$} 
\, ,
\end{array} \right.
\label{eqnapx}
\\ \nonumber
\end{eqnarray}
where
\begin{eqnarray}
N_j(\tau) &\equiv& \sum_\nu \langle T_\tau \, \hat{n}_{j\nu}(\tau) \, 
\hat{n}_{j\nu}(0) \rangle 
\label{eqncapn} \\
D_{jj}(\tau) &\equiv&  \frac12 \sum_{\nu\nu^\prime}^{\nu\neq\nu^\prime}
\langle T_\tau \, \hat{n}_{j\nu}(\tau) \,\hat{n}_{j\nu^\prime}(0) \rangle
\nonumber \\
D_{j\neq j^\prime}(\tau) &\equiv&  \frac12 
\sum_{\nu\nu^\prime}
\langle T_\tau \,
[\hat{n}_{j\nu}(\tau) \,\hat{n}_{j^\prime\nu^\prime}(0)
\nonumber \\
&&\;\;\;\;\;\;\; +\, \hat{n}_{j^\prime\nu^\prime}(\tau) \,\hat{n}_{j\nu}(0)] 
\rangle
\label{eqncapd}
\end{eqnarray}
with $\nu$ ranging over the $2j\!+\!1$ states of $j$ (similarly
$\nu^\prime$ and $j^\prime$).  Then
\begin{equation}
\chi_J(\tau)= 
\sum_j (j\!+\!1)[jN_j(\tau)-D_{jj}(\tau)] \, .
\end{equation}
Note that at $\tau\!=\!0$, $N_j(0)\!=\!n_j$ and
$D_{jj^\prime}(0)\!=\!d_{jj^\prime}$ where the number of electrons
in the $l$-shell is $n_l=n_1\!+\!n_2$ and the associated double
occupation is $d_l=d_{11}\!+\!d_{12}\!+\!d_{22}$, with subscripts $1$
and $2$ labelling $j=l\!-\!1/2$ and $j=l\!+\!1/2$ for $l\!\geq\!1$,
respectively.

\begin{figure}[t]
 \includegraphics[width=3.0in]{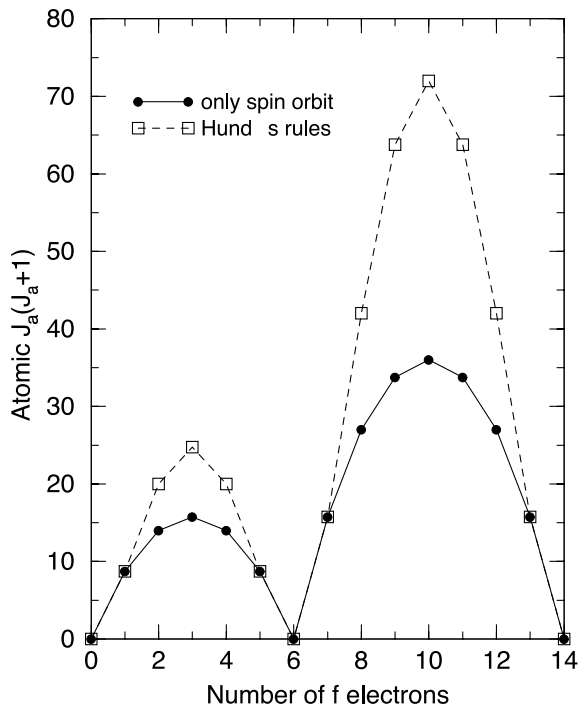}
\caption{Atomic moments $\langle {\bf \hat{J}}^2 \rangle
=J_a(J_a\!+\!1)$ for the rare earths.
\label{figatm}}
\end{figure}

The {\it bare }or instantaneous local moments corresponding to
each of Eqs.~(\ref{chiS}--\ref{chiJ}) are given by their
$\tau\!=\!0$ values, e.g.,
\begin{eqnarray}
J_b(J_b\!+\!1)&=&\langle {\bf \hat{J}}^2 \rangle = \chi_J(\tau\!=\!0)
\nonumber \\
&=& \sum_j (j\!+\!1)[jn_j-d_{jj}] \, .
\label{eqnJb}
\end{eqnarray}
For $f$ electrons with integer $n_f\!=\!n$ shell populations
$0\leq n \leq 14$, one might expect in the strongly localized,
atomic limit that
\begin{eqnarray}
n_1 &=& \min(n,6) \nonumber \\
d_{11} &=& n_1(n_1\!-\!1)/2 \nonumber \\
n_2 &=& \max(0,n-6) \nonumber \\
d_{22} &=& n_2(n_2\!-\!1)/2 \, , \nonumber
\end{eqnarray}
which leads to the filled data points (only spin orbit) in Fig.~1.
Inclusion of the appropriate intraatomic exchange terms would give
the correct Hund's rules values designated by the open squares.
The spin-orbit-only results are seen to give the correct qualitative
behavior with filling, and the correct values of $\langle {\bf
\hat{J}}^2 \rangle$ for subshells with one or no holes or electrons.
As will be seen in this paper, they also appear to give the
qualitatively correct evolution from localized to itinerant behavior
with compression, since much of that originates from volume-dependent
changes in the double occupation which is captured correctly.

\begin{figure}[t]
 \includegraphics[width=3.0in]{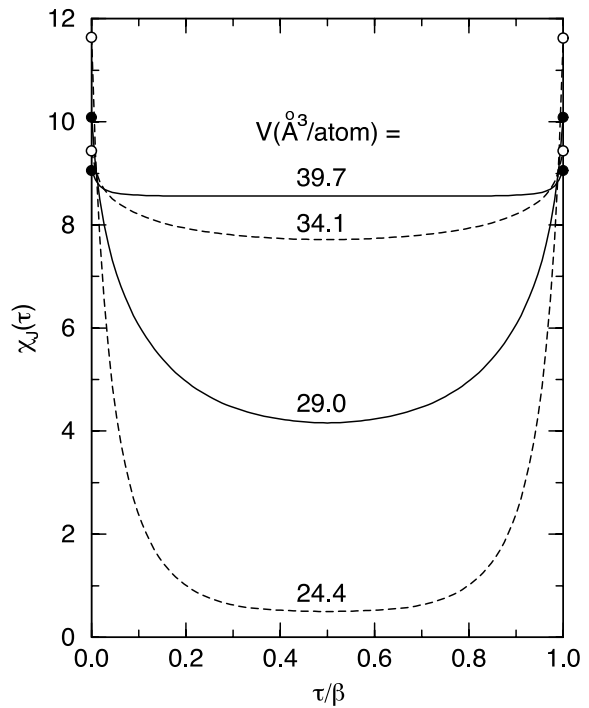}
\caption{ $\chi_J(\tau)$ for Ce at $632\,$K and various volumes. The
$\tau\!=\!0$ values (circles) give the bare moment squared $\langle
\hat{\bf J}^2\rangle = J_b(J_b\!+\!1)$, while the $\tau$ average gives
the screened quantity $J_s(J_s\!+\!1)$.  There is little screening at
large volume, while the two moments are quite different at small volume.
The $\alpha$ and $\gamma$ sides of the collapse are at volumes of $27.8$
and $33.1\,$\AA$^3$/atom, respectively, correspondingly roughly to the
middle two curves.  The equilibrium volume is $34.4\,$\AA$^3$/atom.
\label{figJtau}}
\end{figure}

Information about {\it screened} moments comes from the
static susceptibility $\chi_J(\omega\!=\!0)$.  Given
Curie-Weiss behavior, an effective moment can be extracted
from the slope of $\chi_J(\omega\!=\!0)$ versus $T^{-1}$, thus
\begin{equation}
J_s(J_s\!+\!1)= T\chi_J(\omega\!=\!0) = \frac{1}{\beta}
\int_0^\beta d\tau \chi_J(\tau)
\end{equation}
At large volume and thus weak hybridization, $\hat{\bf J}$ approximately
commutes with the Hamiltonian so that $\chi_J(\tau)\sim$ constant, and
thus there is no screening ($J_s \sim J_b$).  There is also no screening
in the high temperature limit since $\chi_J(\tau)\rightarrow \chi_J(0)$
as $0\!\leq \!\tau \!\leq\!  1/T \rightarrow 0$.  At small volume (strong
hybrization) and low temperature, $\chi_J(\tau)$ falls away between
its maximal values at $\tau=0$ and $\beta$, leading to $J_s < J_b$.
See Fig.~\ref{figJtau}.

\subsection*{Charge susceptibility $\chi_c(\tau)$}

The exact expression for the local charge susceptibility
Eq.~(\ref{eqnchic}) may be written using the definitions
Eqs.~(\ref{eqncapn},\ref{eqncapd}) as
\begin{equation}
\chi_c(\tau) = \sum_j N_j(\tau) + 2\sum_{j\leq j^\prime} 
D_{jj^\prime}(\tau) - n_l^2 \, .
\end{equation}
Following the language of Ref.~\onlinecite{Ylvisaker09} the local charge
``fluctuations'' are
\begin{eqnarray}
\langle \delta \hat{n}_l^2 \rangle  &\equiv&
\langle (\hat{n}_l\!-\!n_l)^2 \rangle 
= \chi_c(\tau\!=\!0) 
\nonumber \\
&=& 2d_l - n_l(n_l\!-1) \, ,
\label{eqnchic0}
\end{eqnarray}
while the local {\it static} charge ``susceptibility'' is
\begin{equation}
\chi_c^{(1)} \equiv \chi_c(\omega\!=\!0) 
= \int_0^\beta d \tau \chi_c(\tau) \, .
\label{eqnchic1}
\end{equation}
Note that $T\chi_c^{(1)}\leq \langle \delta \hat{n}_l^2 \rangle$ since
$T\chi_c^{(1)}$ is the $\tau$ average of $\chi_c(\tau)$ which drops from
its $\tau\!=\!0$, $\beta$ maxima of $\langle \delta \hat{n}_l^2 \rangle$
to smaller values in the mid $\tau$ range, similar to Fig.~\ref{figJtau}
for the spin case.  This upper bound for $T\chi_c^{(1)}$ is of interest
since it helps to identify a large $\chi_c^{(1)}$, which signals the
existence of prominent low-energy charge excitations as occurs, e.g.,
in the Yb valence transition.\cite{Ylvisaker09}

Clearly $\langle \delta \hat{n}_l^2 \rangle$ and $T\chi_c^{(1)}$ are the
charge susceptibility analogs of $J_b(J_b\!+\!1)$ and $J_s(J_s\!+\!1)$,
respectively, for the spin case.  Similarly, $\langle \delta \hat{n}_l^2
\rangle$ and $T\chi_c^{(1)}$ must approach one another in the large
volume localized limit as $\hat{n}_l$ becomes an eigenoperator of the
system with vanishing hybridization.  However, in contrast to the spin
case, $\langle \delta \hat{n}_f^2 \rangle\rightarrow 0$ (and thus also
$\chi_c^{(1)}\rightarrow 0$) in this limit for the trivalent rare earths
since $n_f\rightarrow n$ and $d_f\rightarrow n(n\!-\!1)/2$, where $n$
is the nominal integer $4f$ occupation, e.g., $n\!=\!1$ for Ce.

\section{Configuration weights}

The probabilities or configuration weights $w_k$ of finding integer
$k$ $l$-shell electrons on a given site are useful in discussing
delocalization, and are related to $\chi_c(\tau\!=\!0)=\langle \delta
\hat{n}_l^2 \rangle$ insofar as they may also be expressed as linear
combinations of $n_l$ and $d_l$ near the localized limit.  The $w_k$
are given by
\begin{eqnarray}
w_k &=& Z_k / \sum_{k^\prime} Z_{k^\prime}
\nonumber \\
Z_k &=& \sum_{s_k} \langle ks_k| e^{-\beta(\hat{H}\!-\!\mu \hat{N})}
|ks_k\rangle \, .
\end{eqnarray}
Here $\hat{H}$ is the Hamiltonian; $\hat{N}$, the total number
operator for all types of electrons; $\mu$, the chemical potential;
and $\{|ks_k\rangle\}$, a complete set of eigenstates of $\hat{n}_l$,
$\hat{n}_l|ks_k\rangle = k|ks_k\rangle$, where all other quantum
numbers besides $k$ are lumped into $s_k$.  Evaluating the thermal
expectations $\langle\cdots\rangle$ of $1$, $\hat{n}_l$, and
$\hat{n}_l(\hat{n}_l\!-\!1)/2$ using the same complete basis yields 
\begin{eqnarray}
1 &=& \sum_k w_k
\nonumber \\
n_l &=& \sum_k k \, w_k
\nonumber \\
d_l &=& \sum_k k(k\!-\!1)\, w_k/2 \, ,
\label{eqnstatw}
\end{eqnarray}
which shows the statistical nature of $n_l$ and $d_l$.

At sufficiently large volumes and low temperatures where only $w_k$ for
$k=n$, $n\!\pm\!1$ are non-negligible, these three $w_k$ may be expressed
via Eq.~(\ref{eqnstatw}) in terms of $n_l$ and $d_l$
\begin{eqnarray}
w_{n\!-\!1} &=& d_l - d_l^{\rm min}(n_l) + (|n_l\!-\!n|-n_l+n)/2
\nonumber \\
w_n &=& 1 - 2[d_l\!-\!d_l^{\rm min}(n_l)]  - |n_l\!-\!n|
\nonumber \\
w_{n\!+\!1} &=& d_l - d_l^{\rm min}(n_l) + (|n_l\!-\!n|+n_l-n)/2 \, .
\label{eqnwts}
\end{eqnarray}
Here it is convenient to use a function $d_l^{\rm min}(n_l)$ which is
the minimum possible double occupation for an ensemble of sites whose
average $l$-shell population is $n_l$.  This is a piecewise linear
function which assumes the values $k(k\!-\!1)/2$ at integer $k$ values
of $n_l$, and may be expressed.
\begin{equation}
d_l^{\rm min}(n_l) = (n(n\!-\!1)+(2n\!-\!1)(n_l\!-\!n)+|n_l\!-\!n|)/2
\end{equation}
for the range $n\!-\!1 \leq n_l \leq n\!+\!1$.

Equation~(\ref{eqnwts}) appears intuitively to separate the effects
of delocalization from those arising more simply out of changes in
$n_l$ due to a possible $l$-shell electron reservoir.  Should $n_l$
increase due to such a reservoir while the system is still in
the strongly localized limit, where presumably  $d_l\!=\!d_l^{\rm
min}(n_l)$, then $w_{n\!-\!1}\!=\!0$, $w_n\!=\!1\!-\!n_l\!+\!n$,
$w_{n\!+\!1}\!=\!n_l\!-\!n$.  This suggests $w_{n\!-\!1}$ is untainted
by such reservoir effects for $n_l\!\geq\! n$, or more generally from
Eq.~(\ref{eqnwts})
\begin{eqnarray}
d_l\!-\!d_l^{\rm min}(n_l) =
\left\{ \begin{array}{ll} 
w_{n\!+\!1} &  \mbox{$n\!-\!1\leq n_l \leq n$}  \\
w_{n\!-\!1} &  \mbox{$n\leq n_l \leq n\!+\!1$}  \\
\end{array} \right. \, ,
\label{eqndeloc}
\end{eqnarray}
which we argue in Sec.~V is a useful diagnostic for delocalization.
By contrast the local charge fluctuations
\begin{equation}
\langle \delta \hat{n}_l^2 \rangle =  w_{n\!-\!1}+w_{n\!+\!1}
 -(n_l\!-\!n)^2 
\label{eqncfvswk}
\end{equation}
appear to mix delocalization and reservoir effects and so are therefore
less useful.


\section{Computational details}

The LDA+DMFT calculations reported in this paper have generally
been carried out as in previous work on the compressed rare
earths.\cite{Held01,McMahan03,McMahan05}  All calculations were performed
for an assumed fcc structure, and at a temperature of $632\,$K (4 mRy).
The spin-orbit interaction was included in addition to the scalar part
of the $4f$ Coulomb interaction, i.e., the screened Slater integral
$F^{0}\equiv U_f$, however, not the higher Slater integrals ($F^k$,
$k\!=\!2,4,6$) which describe the Hund's rules intraatomic exchange.
While this gives the wrong values for some $4f$ moments in the localized
limit, the volume-dependence accompanying delocalization of these
moments may still be reasonably captured as this appears to follow from
fairly general behavior in the evolution of such quantities as the
double occupancy.  As in the earlier papers, the LDA contribution to
the present work was provided by linear muffin-tin orbital calculations
in the atomic-sphere approximation as described elsewhere.\cite{JCAMD}
The auxiliary Anderson impurity problem was solved using the Hirsch-Fye
quantum Monte Carlo (QMC) alogorithm,\cite{Hirsch86,QMC1bnd} with
results obtained for $L\!=\!80$ and $112$ time slices extrapolated to
$L\!=\!\infty$ assuming a $1/L^2$ dependence.  This was unnecessary for
$\chi_J(\tau)$ where the two $L$ values gave essential agreement.  The
disposition of $U_f\hat{n}_f$ terms between the kinetic and interaction
parts of the auxiliary impurity Hamiltonian in the Hirsch-Fye QMC,
and the impact of this choice on Trotter corrections, is discussed in
the Appendix.

The susceptibilities reported in this work were calculated within the
QMC using Wick's theorem, e.g., Eq.~(154) of Ref.~\onlinecite{DMFT2}.
They were obtained from runs of $\sim\!350,000$ and $\sim\!100,000$
sweeps for $L=80$ and $112$, respectively, using previously converged
self-energies to get the input bath Green functions.  Error estimates
were obtained from Eq.~(5.3) of Ref.~\onlinecite{Jarrell96} in conjunction
with an examination of the bin-dependence of the data stored as a function
of sweep.



\begin{figure}[t]
\includegraphics[width=3.0in,height=4.2in]{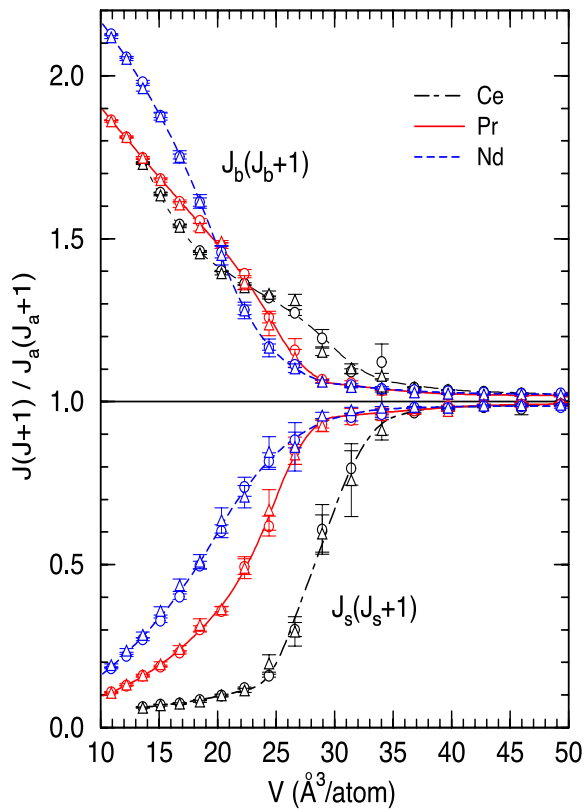}
\caption{(color online). Bare $J_b(J_b\!+\!1)=\chi_J(\tau\!=\!0)$ and
screened $J_s(J_s\!+\!1)=T\chi_J(\omega\!=\!0)$ moments for Ce, Pr, and
Nd at 632 K, relative to the atomic limiting values $J_a(J_a\!+\!1)$ of
Fig.~\ref{figatm} for the ``only spin orbit'' case.  Smoothed curves are
drawn through the points giving $L=80$ (circles) and $112$ (triangles)
results.
\label{figJall}}
\end{figure}

\section{Results}

We now present results of the susceptibility and configuration weight
calculations giving first comparisons between Ce, Pr, and Nd as a function
of volume, then turning to insights provided by these results in regard to
the experimentally observed transitions.  Definitions of the quantities
calculated have been presented in Secs.~II and III.  The volume range
studied is from $10$ to $50\,$\AA$^3$/atom which may be compared to 300-K
experimental volumes of $14.8$, $14.1$, and $14.2\,$\AA$^3$/atom at a
pressure of $100\,$GPa; and $34.4$, $34.5$, and $34.2\,$\AA$^3$/atom at
$0\,$GPa; for Ce,\cite{VohraCe99,young91} Pr,\cite{ChesnutPr00,young91}
and Nd,\cite{ChesnutNd00,young91} respectively.

Figure \ref{figJall} shows results for $J(J\!+\!1)$ corresponding
to the bare $J_b(J_b\!+\!1)=\chi_J(\tau\!=\!0)$ and screened
$J_s(J_s\!+\!1)=T\chi_J(\omega\!=\!0)$ moments in Ce, Pr, and Nd, divided by
the atomic limiting values $J_a(J_a\!+\!1)=8.75$, $14$, and $15.75$,
respectively, from Fig.~\ref{figatm}.  These are the ``only spin orbit''
values of that figure, which give the correct $J(J\!+\!1)$ for Ce,
however, are 30 and 36\% smaller than the true Hund's rules values for Pr
and Nd, respectively. It is particularly evident for the screened results
that the changes are most abrupt and occur at the largest volumes for Ce,
and then successively moderate and shift to smaller volume for Pr and
then Nd, a pattern which will be seen throughout the present results.
The bare moments increase with compression simply because the $n_f$
values increase.\cite{nfvsV} Since a completely random population of the
$4f$ states would also have $\langle \hat{J}_z^2\rangle$ and therefore
$\langle \hat{\bf{J}}^2\rangle$ increase with $n_f$, this does not imply
the kind of coherent physical moment at the smallest volumes that one
certainly has in the large volume localized limit.

\begin{figure}[t]
\includegraphics[width=3.0in,height=4.2in]{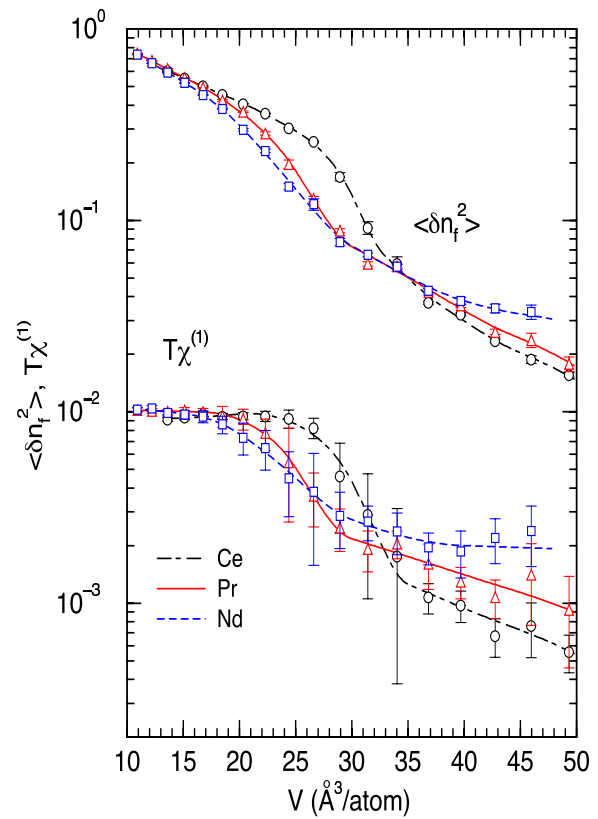}
\caption{(color online).  Local charge fluctuations $\langle \delta
\hat{n}_f^2 \rangle = \chi_c(\tau\!=\!0)$ and the static charge
susceptibility $\chi_c^{(1)} = \chi_c(\omega\!=\!0)$, the latter
multiplied by $T$, for Ce, Pr, and Nd at 632 K.  Both show rapid
increases under compression coincident with the changes observed
in Fig.~\ref{figJall}.  $T\chi_c^{(1)}$ is everywhere more than an
order of magnitude smaller than $\langle \delta \hat{n}_f^2 \rangle$
reflecting the absence of low-energy charge fluctuations, and the
ratio $T\chi_c^{(1)}/\langle \delta \hat{n}_f^2 \rangle$ becomes even
smaller at small volumes reflecting screening of the charge fluctuations.
All points are $1/L^2$ extrapolations to $L\!=\!\infty$, with smoothed
curves drawn through them.
\label{figchg}}
\end{figure}

Figure~\ref{figchg} shows the local charge fluctuations
$\langle\delta\hat{n}_f^2\rangle=\chi_c(\tau\!=\!0)$ and $T$
times the local static charge susceptibility $T\chi_c^{(1)} =
T\chi_c(\omega\!=\!0)$, which are the charge analogs of the bare and
screened $J(J\!+\!1)$ of Fig.~\ref{figJall}.  Both quantities show
rapid increases with compression coincident with the changes seen in
Fig.~\ref{figJall}.  Like the spin case these two quantities must also
approach one another in the large volume limit at low temperatures,
although these limiting values are not finite but $0$ for the charge
case.  This follows from $n_f\!\rightarrow \!n$ and $d_f\!\rightarrow\!
n(n\!-\!1)/2$ in this limit, where $n\!=\!1$ (Ce), $2$ (Pr), and $3$ (Nd),
and given that $\langle\delta\hat{n}_f^2\rangle=2d_f-n_f(n_f\!-\!1)$
from Eq.~(\ref{eqnchic0}) and $\langle\delta\hat{n}_f^2\rangle\geq
T\chi_c^{(1)}$ as discussed in Sec.~II.  The significance of the vanishing
$4f$ charge fluctuations in the large-volume, low-temperature limit is
disappearance of the Kondo resonance thus leaving a gapped $4f$ spectra
overlapping the Fermi level.

Also interesting in Fig.~\ref{figchg} is how much smaller the
local static charge susceptibility $\chi_c^{(1)}$ is compared
to its limiting maximum $\langle\delta\hat{n}_f^2\rangle/T$,
e.g., $T\chi_c^{(1)}/\langle\delta\hat{n}_f^2\rangle\!=\!0.017$,
$0.031$, and $0.033$ at $V\!=\!15$, $28$, and $41\,$\AA$^3$/atom,
respectively, for Ce, with similarly small values for Pr and Nd.
In contrast this ratio is $0.25$, $0.58$, and $0.09$ at the same
volumes for the second to the last rare earth, Yb.\cite{Ylvisaker09}
As pointed out in Ref.~\onlinecite{Ylvisaker09} the small ratios for
Ce, Pr, and Nd consitute normal behavior where the large onsite Coulomb
interaction supresses low-energy charge excitations leading to a small
local static susceptibility $\chi_c^{(1)}$.  The oddball is Yb which
has a valence transition from divalent ($f^{14}$) at large volume
to trivalent ($f^{13}$) at small volume.  The large Yb ratio $0.58$
at $28\,$\AA$^3$/atom is in the midst of the valence transition where
the near degeneracy of $f^{13}$ and $f^{14}$ configurations leads to
prominent low-energy charge excitations and a consequent large local
static charge susceptibility.  Finally, all four rare earths exhibit
decreasing ratios $T\chi_c^{(1)}/\langle\delta\hat{n}_f^2\rangle$ for
compression approaching the smallest volumes considered, which reflects
screening of the charge fluctuations similar to screening of the moments
as has been noted.\cite{Ylvisaker09}

\begin{figure}[t]
\includegraphics[width=3.0in,height=4.2in]{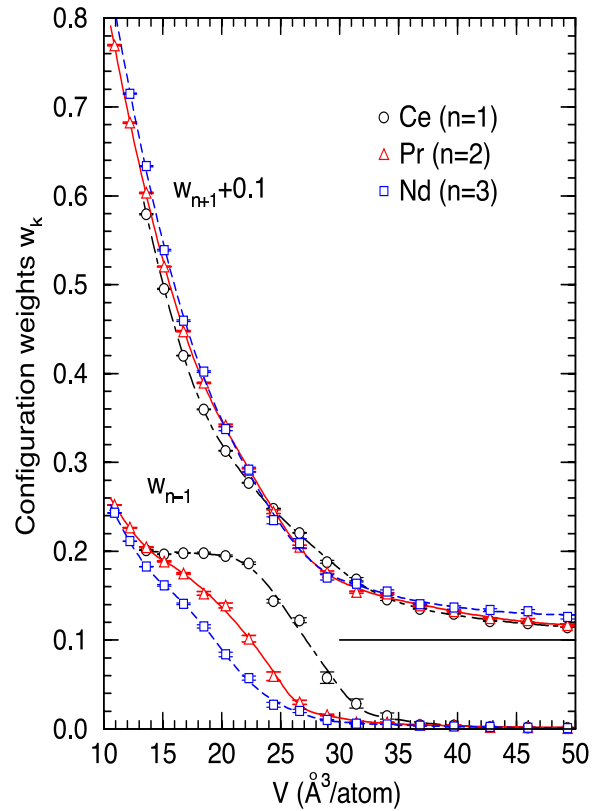}
\caption{(color online).  Configuration weights $w_{n\!-\!1}$ and
$w_{n\!+\!1}$ (latter $+0.1$) for Ce $(n\!=\!1)$, Pr $(n\!=\!2)$,
and Nd $(n\!=\!3)$.  The former is a measure of delocalization
$w_{n\!-\!1}=d_f\!-d_f^{\rm min}(n_f)$, since $n\lesssim n_f < n\!+\!1$
everywhere, and shows a dramatic rise with compression for Ce, structure
which successively softens and shifts to smaller volumes in moving to Pr
and then Nd. The weights $w_{n\!+\!1}$ are complicated by the general
increase in $n_f$ with compression due to the rising $6s,6p$ bands.
All points are $1/L^2$ extrapolations to $L\!=\!\infty$, with smoothed
curves drawn through them.
\label{figwk}}
\end{figure}

The probability or configuration weight $w_k$ of finding integer $k$
$f$ electrons on a given site is of some interest in understanding the
manner in which the rare earths evolve from localized to itinerant
character under compression.  As all sites are pinned at specific
integer occupations $n$ in the large-volume, localized limit at
low temperatures, then for some range of smaller volumes away
from this limit only $w_n$ and $w_{n\pm1}$ are nonneglibile, and may be
determined from the average number of $4f$ electrons $n_f$ and their
associated double occupation $d_f$ according to Eq.~(\ref{eqnwts}).
Figure~\ref{figwk} shows $w_{n\!-\!1}$ and $w_{n\!+\!1}$ (the latter
$+0.1$ for visual clarity) calculated in this manner for Ce, Pr, and Nd
($n\!=\!1$, $2$, and $3$, respectively).  Equation~(\ref{eqnwts}) assumes
$w_n = 1\!-\!w_{n\!-\!1}\!-\!w_{n\!+\!1}$, so that the limit $w_n\!=\!1$,
$w_{k\neq n}\!=\!0$ is evidently  being approached at large volumes.

As volume is reduced away from the localized limit, two things happen.
First, for purely one-electron reasons, the $6s$ and $6p$ bands rise
relative to the $4f$ levels, causing an increase in $n_f$,\cite{nfvsV}
and thus a shift in weight from $w_n$ to $w_{n\!+\!1}$.  Second, as
hybridization grows, the $4f$ electrons begin to hop to $f$ states on
neighboring sites or into and out of valence levels, causing {\it both}
$w_{n\pm 1}$ to grow at the expense of $w_n$.  Only $w_{n\!-\!1}$ is a
true measure of the second, delocalization effect uncomplicated by the
consequences of increasing $n_f$ \cite{footnote}.  
Or more generally, this diagnostic for
delocalization is given by Eq.~(\ref{eqndeloc}) which is $w_{n\!-\!1}$
for $n_f\geq n$ and $w_{n\!+\!1}$ for $n_f\leq n$.

Figure \ref{figwk} shows clear evidence of both the the rising $6s$,
$6p$ bands and delocalization.  Note first the onset of delocalization
as $w_{n\!-\!1}$ increases with compression, and how much more abrupt
this behavior is compared to the smoother $w_{n\!+\!1}$ curves which
reflect also the effects of increasing $n_f$.  These onsets occur near
the equilibrium volume of Ce ($V_0\!=\!34.37\,$\AA$^3$/atom), but at
somewhat smaller volumes than the corresponding $V_0$ of Pr ($34.54$)
and Nd ($34.17$).  At the largest volumes ($V\!>\!42\,$\AA$^3$/atom),
$w_{n\!+\!1}$ is an order of magnitude larger than $w_{n\!-\!1}\leq 0.003$
for all three rare earths, suggesting the behavior there is predominantly
the increase in $n_f$ due to the rising $6s$, $6p$ bands.  The Fermi
level in this regime lies midway between the two Hubbard bands, and a
DMFT calculation with the Hubbard I self-energy---which is incapable of
generating the quasiparticle peak---has $n_f$ pinned to 1 in the case
of Ce for all volumes $V\!>\!17.6\,$\AA$^3$/atom.  This suggests that
the large-volume shift in weight $w_n\!\rightarrow\!w_{n\!+\!1}$ is
not mixed valence in the sense of the Fermi level moving into the upper
Hubbard band, but rather that the rising $6s$ and $6p$ bands transfer
electrons into the $4f$ quasiparticle peak, which is nonetheless still
quite small in this region.

There have been experimental determinations of the Ce
configuration weights $w_{0,1,2}$ using Anderson impurity model
analyses of photoemission\cite{KVC3} and resonant inelastic
x-ray scattering\cite{Rueff06} data, and a number of LDA+DMFT
calculations\cite{Zoelfl01,Held01,McMahan03,McMahan05,Amadon06} of these
quantities or related $n_f = 1\!-\!w_0\!+\!w_2$.  All are consistent with
$w_1\!\rightarrow\! 1$, $w_{k\neq1}\!\rightarrow \!0$ in the large volume
limit.  There is an asymmetry between $f^1$-$f^0$ and $f^1$-$f^2$ mixing
in solutions of the Anderson impurity model which tends to lead to the
predominant transfer $w_1\!\rightarrow\! w_0$ with growing hybridization,
and thus a decrease in $n_f$ across the $\gamma$ to $\alpha$ collapse
and $w_0>w_2$ in the $\alpha$ phase.\cite{KVC4}   This behavior is
seen in the analyses of both experimental papers.\cite{KVC3,Rueff06}
The present and our earlier work\cite{Held01,McMahan03,McMahan05} concur
with the predominant transfer $w_1\!\rightarrow\!w_0$ across the
collapse, however, has everywhere a larger $w_2$ than these analyses,
quite possibly reflecting the impact on $n_f$ of the rising $6s$, $6p$
bands which may not be well treated in the impurity model simulations.
Thus we see a non-monotonic dip in $n_f$ across the collapse, but generally
$n_f>1$ and $w_0< w_2$.\cite{nfvsV} 

\begin{figure}[t]
\includegraphics[width=3.0in,height=4.4in]{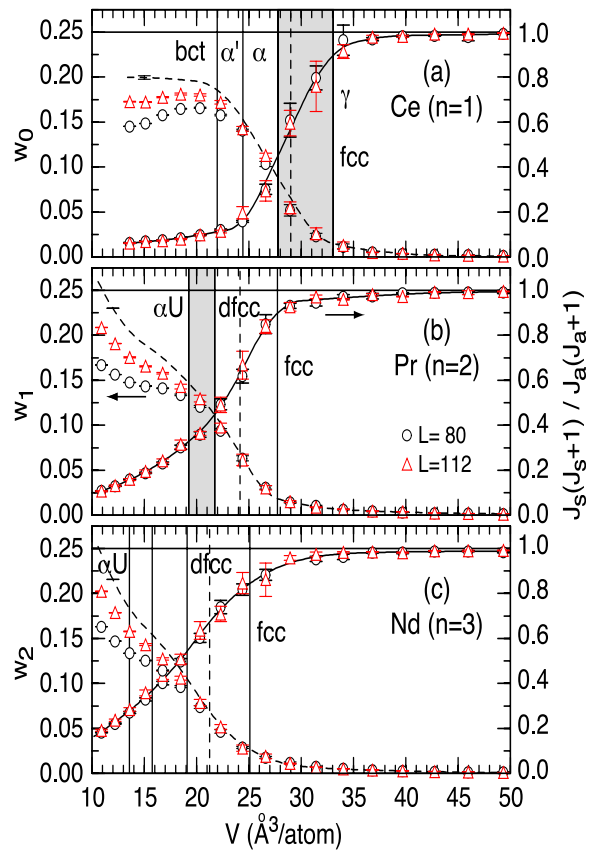}
\caption{(color online).  Screened $J_s(J_s\!+\!1)=T\chi_J(\omega\!=\!0)$
moments (solid curves) and configuration weights $w_{n\!-\!1}$ (dashed
curves) for (a) Ce, (b) Pr, and (c) Nd at 632 K.  The former are divided
by the atomic limiting values $J_a(J_a\!+\!1)$ of Fig.~\ref{figatm} for
the ``only spin orbit'' case.  The smoothed curves are drawn through the
combined $L=80$ (circles) and $112$ (triangles) points for the moments,
and extrapolated to $L\!\rightarrow\!\infty$ for the $w_{n\!-\!1}$.
All calculations were for an assumed fcc structure, however, the results
are compared to the observed phases demarcated by the vertical lines,
with shading for Ce and Pr indicating significant two phase regions.
The vertical dashed lines show the volumes where the Kondo temperature
is 632 K.
\label{figexpstr}}
\end{figure}

We turn now to possible insights provided by the present results into
the experimentally observed volume collapse transitions (shaded regions)
and phase structure (demarcated by vertical solid lines) as shown in
Fig.~\ref{figexpstr} for Ce,\cite{OlsenCe85,McMahonCe97,VohraCe99}
Pr,\cite{ZhaoPr95,ChesnutPr00,BaerPr03,CunninghamPr05} and
Nd.\cite{AkellaNd99,ChesnutNd00} At large volume only the face centered
cubic (fcc) phase is labelled, which is the end of the localized,
trivalent rare earth series,\cite{Duthie77} with all three rare earths
found in the preceeding, double-hexagonal close packed phase at ambient
conditions.  The two most dramatic theoretical diagnostics are shown
for comparison, $J_s(J_s\!+\!1)$ and $w_{n\!-\!1}$, and while they were
calculated everywhere assuming an fcc structure, it is hoped nonetheless
that their volume dependence is sufficiently insensitive to structure so
as to still provide useful insights.  This is not an issue for Ce where
both $\alpha$ and $\gamma$ phases bounding the volume collapse are fcc.

LDA+DMFT
calculations\cite{Zoelfl01,Held01,McMahan03,Haule05,McMahan05,Amadon06}
have consistently supported the KVC scenario\cite{KVC1,KVC2,KVC3,KVC4}
for Ce.  They show, e.g., a rapid build up of the Kondo resonance for
volume reduced across the $\gamma$-$\alpha$ two-phase region, which is
taken as a signature of the onset of screening.  The present results
now directly report the screened moment, and indeed one sees a 60\%
reduction in $J_s(J_s\!+\!1)$ from the $\gamma$ to the $\alpha$ side of
the collapse.  Moreover, if the Kondo temperature $T_{\rm K}$ is defined
by the condition $T\chi_J(\omega\!=\!0)/\chi_J(\tau\!=\!0)=0.5$, then
$T_{\rm K}\!=\!632\,$K at the volume $V\!=\!29.0\,$\AA$^3$/atom (vertical
dashed line) for the isothermal results in Fig.~\ref{figexpstr}(a).
This value is consistent with the combined experimental results of
Refs.~\onlinecite{Rueff06} and \onlinecite{KVC3}, except for the middle
10-kbar point of the former which appears out of place given the roughly
exponential behavior expected for $T_{\rm K}(V)$.\cite{KVC1}

The competing MT scenario for the Ce volume collapse relies
on $4f$ electron delocalization as the underlying driving
mechanism.\cite{Johansson74,Johansson75} If Eq.~(\ref{eqndeloc}) is
accepted as one suitable diagnostic, then $w_0$ in Fig.~\ref{figexpstr}(a)
suggests that the collapse also coincides with delocalization, a point
that has been made previously.\cite{Held01,McMahan03} To be careful, note
that Eq.~(\ref{eqnwts}) for the $w_k$ does break down when configurations
$f^k$ for $|k\!-\!n|\geq 2$ become important as the itinerant limit
is approached.  It would seem from the present and earlier LDA+DMFT
calculations that the following are all different facets of the same
continuous and extended evolution with compression of a nominally
$f^n$ trivalent rare earth: (1) transfer of $4f$ spectral weight from
Hubbard side bands to a Fermi-level structure, (2) screening of the $4f$
moments, (3) growth of $4f$ charge fluctuations, and (4) dispersal of
configuration weight away from $w_n\!=\!1$ to neighboring and then more
distant configurations.  Some of these facets look more Kondo like,
and some seem more consistent with intuitive ideas of delocalization.

There is striking similarity between the theoretical diagnostics for Ce
and those for Pr and Nd in Fig.~\ref{figexpstr}, although the screening
and delocalization effects shift to smaller volume and become more gradual
for Pr, and then more so for Nd.  While the Ce evolution seems to have
saturated under compression on reaching the body centered tegragonal (bct)
phase at about $V\!=\!22\,$\AA$^3$/atom ($P\!=\!13\,$GPa),\cite{OlsenCe85}
Nd has yet to reach this stage even by the smallest volume considered
$10\,$\AA$^3$/atom ($P\!\sim\!365\,$GPa),\cite{Nd10} and Pr is
intermediate.  
The stability fields of the distorted fcc (dfcc) structures in both
Pr and Nd appear analogous to the $\alpha$-$\gamma$
two phase region in Ce, according to progression of the two
theoretical diagnostics.  Only on further compression is there a volume
collapse in Pr from dfcc to $\alpha$-U, while Nd passes
through two additional phases before reaching the same $\alpha$-U structure,
absent any 
large volume changes.
It is conceivable that this progression in
collapse size and location follows simply from the shift to smaller
volumes and moderation in the correlation contributions which must then
compete with the ever bigger underlying benign part of the equation
of state.  While there are suggestions of a Van der Waals loop in
the LDA+DMFT free energy corresponding to the isostructural fcc Ce
collapse,\cite{Held01,McMahan03,McMahan05,Amadon06} one must worry
about the need to include the proper Hund's rules exchange for multi-$f$
electron Pr and Nd, as well as performing the calculations for all of
the observed structures, e.g. $\alpha$-U.  Moreover, the Ce collapse
has a critical temperature of only $480\,$K,\cite{Lipp08} while for
Pr, the dfcc phase is absent above about $700\,$K with yet a new Pr-VI
phase intermediate between the $\alpha$-U and fcc phases.\cite{BaerPr03}
These temperature senstivities are also a reminder of the need to include
lattice vibrational contributions, which may themselves further modify
the nature of the collapse transitions.\cite{Jeong04,Lipp08}

\section{Summary}

We have reported LDA+DMFT calculations as a function of volume at
$632\,$K, for the $4f$ spin and charge susceptibilities, and the
probabilities of finding $4f^{n\pm 1}$ configurations in the nominally
$4f^n$ trivalent rare earths Ce ($n\!=\!1$), Pr ($n\!=\!2$), and Nd
($n\!=\!3$).  We find these metals to remain localized at pressures
up through the fcc ($\gamma$-Ce) phases, the last structure of the
initial close-packed series,\cite{Duthie77} as indicated by sharp
$4f^n$ populations, unscreened moments with atomic-limiting values,
and small charge fluctuations indicating little $4f$ state density
overlapping the Fermi level.  On subsequent compression there is rapid
and dramatic screening of the moments and concurrent increase in charge
fluctuations and the static charge susceptibility.  These changes are
also coincident with rapid growth in the $4f^{n\!-\!1}$ configuration
weights, which we argue offer a truer measure of delocalization than do
the $4f^{n\!+\!1}$ weights which are complicated by the overall increase
in the number of $4f$ electrons due to the rising $6s$, $6p$ bands.
Combined with earlier LDA+DMFT results, this work suggests a continuous
and extended evolution with compression of a nominally $4f^n$ trivalent
rare earth in which there is (1) transfer of $4f$ spectral weight from
Hubbard side bands to the vicinity of the Fermi level, (2) screening
of the $4f$ moments, (3) growth of $4f$ charge fluctuations, and (4)
dispersal of configuration weight away from $4f^n$ to adjacent and then
more distant configurations $4f^k$.  The static charge susceptibility
$\chi_c^{(1)}$ mirrors the volume dependence of the charge fluctuations
$\langle \delta\hat{n}_f^2 \rangle$, and further indicates screening of
these fluctuations at small volume, and supression of low-energy charge
fluctuations by the Coulomb interaction at all volumes.\cite{Ylvisaker09}

The composite evolution in Ce persists until the bct phase
($V\!<\!12\,$\AA$^3$/atom, $P\!>\!13\,$GPa),\cite{OlsenCe85} whereas the
$\alpha$-$\gamma$ two-phase region coincides with only the large-volume
half or so of this progression.  These effects shift to smaller volume,
become more gradual, and extend over a broader volume range for Pr, and
even more so for Nd, which has yet to saturate by $10\,$\AA$^3$/atom
($P\!\sim\!365\,$GPa).\cite{Nd10}  The observed stability fields
of the dfcc phases in Pr and Nd also coincide with the large-volume
half of the evolution in these materials, and thus correspond to the
$\alpha$-$\gamma$ two-phase region in Ce.  The collapse is observed
in Pr only on further compression, and absent in Nd.  Since LDA+DMFT
calculations do hint at a Van der Waals loop in the Ce free energy
at about the right place,\cite{Held01,McMahan03,McMahan05,Amadon06}
the full story for Pr and Nd may await inclusion of the Hund's rules
exchange for these multi-$f$ electron cases, assumption of the correct
structures for all phases (e.g., $\alpha$-U), and possibly also adding
lattice vibrational contributions.\cite{Jeong04,Lipp08}
Even so, this work suggests that the
collapse transitions may not be the predominant story in the compressed
trivalent rare earths, but rather consequences of an underlying and more
robust evolution associated with $4f$-electron delocalization.

\section*{ACKNOWLEDGEMENTS}

Work by AKM was performed under the auspices of the U.S. Department
of Energy by Lawrence Livermore National Laboratory under Contract
DE-AC52-07NA27344.  Work by MJ and RTS was supported by the SciDAC
program, grant DOE-DE-FC0206ER25793.  This work has also benefited from
an alliance with members of the DOE/BES funded Computational Materials
Science Network Cooperative Research Team on ``Predictive Capability
for Strongly Correlated Systems.''  AKM is grateful for conversations
with K. Held, J. Kune\v{s}, W. E. Pickett, and E. R. Ylvisaker.

\section*{APPENDIX: TROTTER CORRECTIONS AND THE BATH GREEN FUNCTION}

One issue related to application of Hirsch-Fye\cite{Hirsch86} QMC
to the rare earth series deserves attention.  To prepare for the
Hubbard-Stratonovich transformation, one rewrites the interaction part
$\hat{I}$ of the auxiliary impurity Hamiltonian
\begin{eqnarray}
U_f \sum_{\alpha < \alpha^\prime} \hat{n}_\alpha \hat{n}_{\alpha^\prime}
&=& U_f \sum_{\alpha < \alpha^\prime} [ \hat{n}_\alpha
\hat{n}_{\alpha^\prime}\!
-\!\mbox{$\frac12$} (\hat{n}_\alpha\!+\!\hat{n}_{\alpha^\prime})]
\nonumber \\
&& + \, \mbox{$\frac{13}{2}$} U_f \hat{n}_f \, ,
\label{eqnHS}
\end{eqnarray}
where $\alpha$ labels the 14 $f$ states.  A strict generalization of
the Hirsch-Fye treatment would be to remove the $U_f \hat{n}_f$ term
from the right side of in Eq.~(\ref{eqnHS}) and add it to the one-body
or kinetic energy part $\hat{K}$ of the Hamiltonian, with the consequent
changes in unperturbed or bath Green function ${\cal G}(\tau)$ and also
the Trotter breakup.  However, for $U_f=6\,$eV and $T=600\,$K, this would
lead to an $f$ bath Green function, absent hybridization and spin orbit
for simplicity, of
\begin{eqnarray}
{\cal G}_f(\tau) &=& -\frac{1}{14}\sum_\alpha
Tr [\hat{\rho} f_\alpha(\tau) f_\alpha^\dagger(0)]/Tr[\hat{\rho}]
\nonumber \\
&\sim &
\frac{-e^{-\tau(\varepsilon_f\!+6.5U_f\!-\!\mu)}}
{e^{-\beta(\varepsilon_f\!+6.5U_f\!-\!\mu)}+1}
\sim - 10^{-300 \tau\!/\!\beta} \, .
\label{eqnbathf}
\end{eqnarray}
Here $\hat{\rho}=exp[-\beta(\hat{K}\!-\!\mu\hat{N})]$, $\mu$ is the
chemical potential, $\hat{N}$ the total electron number operator,
$0<\tau<\beta$, and we take a Ce like site energy $\varepsilon_f\!-\!\mu
\sim -U_f/2$.  Such a function may get into a region of underflow
errors at large $\tau$ in numerical computation.  Note this problem
arises here in the seven-band case only because the $U_f\hat{n}_f$
term in question is 13 times larger than in the familiar one-band case.
For this reason we left the $U_f\hat{n}_f$ term alone in previous work
on the early rare earths,\cite{Held01,McMahan03,McMahan05} and took the
kinetic and interaction parts of the auxiliary impurity Hamiltonian to be
\begin{eqnarray}
\hat{K}&=& \varepsilon_f \hat{n}_f +
\sum_{\lambda, \lambda^\prime} c^\dagger_\lambda
\Delta t_{\lambda\lambda^\prime} c_{\lambda^\prime}
\label{eqnK} \\
\hat{I} &=&  U_f \sum_{\alpha < \alpha^\prime} [ \hat{n}_\alpha
\hat{n}_{\alpha^\prime}\!- \! \mbox{$\frac12$} (\hat{n}_\alpha\!+
\!\hat{n}_{\alpha^\prime})]
+ \, \mbox{$\frac{13}{2}$} U_f \hat{n}_f \, ,
\label{eqnI}
\end{eqnarray}
where $\lambda$ ranges over both $f$ and the additional bath degrees of
freedom, and $\Delta t$ covers the rest of the one-body terms.  Keeping
the $U_f\hat{n}_f$ term in $\hat{I}$ can be handled with a minor
modification of the Hirsch-Fye technique.


\begin{figure}[t]
 \includegraphics[width=3.0in,height=4.2in]{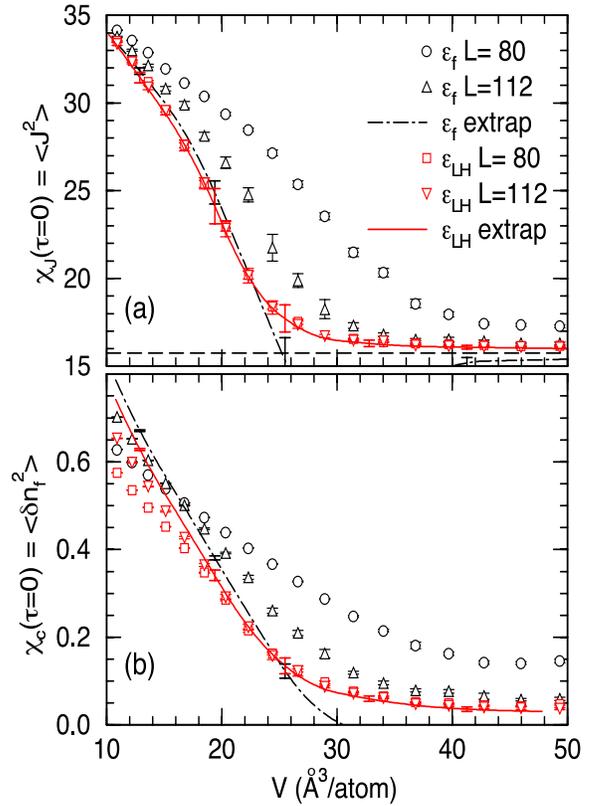}
\caption{(color online).  Trotter corrections in (a)
$\chi_J(\tau\!=\!0)\!=\!\langle {\bf \hat{J}}^2 \rangle$ and (b)
$\chi_c(\tau\!=\!0)\!=\!\langle \delta\hat{n}_f^2 \rangle$ for  Nd as a
function of volume.  Results for $L=80$ and $112$ time slices are shown,
using the $\varepsilon_f$ (circles, up triangles) and $\varepsilon_{\rm
LH}$ (squares, down triangles) effective $f$ site energies in the QMC.
The lines show $L^{-2}$ extrapolation to $L\!=\!\infty$.  It is
evident that the Trotter corrections are considerably smaller using
the $\varepsilon_{\rm LH}$ choice Eqs.~(\ref{eqnKmod},\ref{eqnImod}).
The horizontal dashed line in (a) is the atomic limiting value.
\label{figtrot}}
\end{figure}
Since $\varepsilon_f\!-\!\mu$ itself varies from about $-0.5U_f$ for Ce
to about $-6.5U_f$ for Lu, the choice Eq.~(\ref{eqnK}) then leads to
similar problems for the bath Green function of the late rare earths.
A more consistent treatment for the series as a whole would be to use one
of the Hubbard bands as the effective $f$ site energy in the bath Green
function, and we use the approximate position of the lower Hubbard band,
$\varepsilon_{\rm LH} = \varepsilon_f + (n\!-\!1)U_f$, with the integer
$n$ being the nominal $f$ occupation.  Thus we take
\begin{eqnarray}
\hat{K}&=& [ \varepsilon_f +(n\!-\!1)U_f]\,\hat{n}_f +
\sum_{\lambda,\lambda^\prime} c^\dagger_\lambda
\Delta t_{\lambda\lambda^\prime} c_{\lambda^\prime} \,
\label{eqnKmod}\\
\hat{I} &=& U_f \sum_{\alpha < \alpha^\prime} [ \hat{n}_\alpha
\hat{n}_{\alpha^\prime} - \mbox{$\frac12$} (\hat{n}_\alpha\!+
\!\hat{n}_{\alpha^\prime})]
\nonumber \\
&&+ \, (\mbox{$\frac{15}{2}$}\!-\!n) U_f \hat{n}_f \, ,
\label{eqnImod}
\end{eqnarray}
which was successfully used in Hirsch-Fye calclations for the second
to last rare earth, Yb ($n\!=\!13$),\cite{Ylvisaker09}
while for Ce ($n\!=\!1$), Eqs.~(\ref{eqnKmod},\ref{eqnImod})
are identical to Eqs.~(\ref{eqnK},\ref{eqnI}).

We note here that even for the light rare earths such as Pr ($n\!=\!2$)
and Nd ($n\!=\!3$) that Eqs.~(\ref{eqnKmod},\ref{eqnImod}) serve to
diminish the Trotter corrections relative to Eqs.~(\ref{eqnK},\ref{eqnI}).
This is seen for Nd in Fig.~\ref{figtrot}, where the instantaneous
spin $\chi_J(\tau\!=\!0)\!= \!J_b(J_b\!+\!1)$ and charge
$\chi_c(\tau\!=\!0)\!=\!\langle \delta \hat{n}_f^2 \rangle$
susceptibilities are shown as functions of volume.  Results using
$\varepsilon_f$ [Eqs.~(\ref{eqnK},\ref{eqnI}), open symbols] and
$\varepsilon_{\rm LH}$ [Eqs.~(\ref{eqnKmod},\ref{eqnImod}), closed
symbols] as effective $f$ site energies are given for $L\!=\!80$
and $112$ time slices.Ê The former points are quite different
indicating significant Trotter corrections, while the latter are
essentially on top of one another except for the charge case at the
smallest volumes.  It is reassuring that $L^{-2}$ extrapolations to
$L\!=\!\infty$ (lines) for the $\varepsilon_f$ and $\varepsilon_{\rm
LH}$ cases are fairly close at smaller volumes.  On the other hand,
extrapolation of the $\varepsilon_f$  results yields an unphysical
negative $\chi_c(\tau\!=\!0)$ at large volumes (not shown) indicating
non $L^{-2}$ behavior and the need for larger $L$.  By contrast
the $\varepsilon_{\rm LH}$ results are already converged by $L\!=\!80$
for both spin and charge cases at large volume.  It is clear that
Eqs.~(\ref{eqnKmod},\ref{eqnImod}) represent the superior approach,
and all results in the present paper have been obtained in this manner.



\begin{thebibliography}{99}

\bibitem{Duthie77}
J.C. Duthie, and D.G. Pettifor, Phys. Rev.  Lett. {\bf 38} 564 (1977).


\bibitem{Benedict93}
U. Benedict, J. Alloys Comp. {\bf 193}, 88 (1993).

\bibitem{Holzapfel95}
W. B. Holzapfel, J. Alloys Comp. {\bf 223}, 170 (1995).

\bibitem{JCAMD}
A. K. McMahan, C. Huscroft, R. T. Scalettar, and E. L. Pollock,
J. Comput.-Aided Mater. Design {\bf 5}, 131 (1998).


\bibitem{AkellaNd99}
J. Akella, S. T. Weir, Y. K. Vohra, H. Prokop, S. A. Catledge,
and G. N. Chesntnut, J. Phys.: Condens. Matter {\bf 11}, 6515 (1999).

\bibitem{ChesnutNd00}
G. N. Chesnut and Y. K. Vohra, Phs. Rev. B {\bf 61}, R3768 (2000)

\bibitem{KoskimakiCe78} 
D. G. Koskimaki and K. A. Gschneidner Jr., in {\it Handbook on the
Physics and Chemistry of Rare Earths}, edited by K. A. {Gschneidner
Jr.} and L. R. Eyring (North-Holland, Amsterdam, 1978), p. 337.

\bibitem{OlsenCe85} 
J. S. Olsen, L. Gerward, U. Benedict, J. P. Iti\'e, Physica {\bf
133B}, 129 (1985).

\bibitem{McMahonCe97} 
M. I. McMahon and R. J. Nelmes, Phys. Rev. Lett {\bf 78}, 3884 (1997).

\bibitem{VohraCe99} 
Y. K. Vohra, S. L. Beaver, J. Akella, C. A. Ruddle, and S. T. Weir,
J. Appl. Phys. {\bf 85}, 2451 (1999).

\bibitem{MaoPr81}
H. K. Mao, R. M. Hazen, P. M. Bell, and J. Wittig, J. Appl. Phys.
52, 4572 (1981);
G. S. Smith and J. Akella, J. Appl. Phys. {\bf 53}, 9212 (1982);
W. A.  Grosshans and W. B. Holzapfel, J. Phys.  (Paris) {\bf 45},
C8 (1984).

\bibitem{ZhaoPr95}
Y.C. Zhao, F. Porsch, and W. B. Holzapfel, Phys.  Rev. B {\bf 52},
134 (1995).

\bibitem{ChesnutPr00}
G. N. Chesnut and Y. K. Vohra, Phys. Rev. B {\bf 62}, 2965 (2000).

\bibitem{BaerPr03}
B. J. Baer, H. Cynn, V. Iota, C.-S. Yoo, and G. Shen, Phys. Rev. B
{\bf 67}, 134115 (2003).

\bibitem{CunninghamPr05}
N. C. Cunningham, N. Velisavljevic, and Y. K. Vohra Phys. Rev. {\bf
B} 71, 012108 (2005)

\bibitem{HuaGd98}
H. Hua, Y. K. Vohra, J. Akella, S. T. Weir, R. Ahuja, and
B. Johansson, Rev. High Press. Sci. \& Technol, {\bf 7}, 233 (1998)

\bibitem{PattersonDy04}
R. Patterson, C. K. Saw, and J. Akella, J. Appl. Phys. {\bf 95},
5443 (2004).


\bibitem{mcewen78}
McEwen, K.A., in Gschneidner, Jr., K.A. and Eyring L.R.  (Eds) Handbook
on the Physics and Chemistry of Rare Earths, North-Holland, Amsterdam,
1978, Vol. 1 -- metals, p.~411, see Table 6.1.

\bibitem{ward86}
Ward, J.W., Kleinschmidt, P.D. and Peterson, D.E., in Freeman, A.J. and
Keller, C., (Eds), Handbook on the Physics and Chemistry of the Actinides,
North-Holland, Amsterdam, 1986, Vol. 4, p.~309.

\bibitem{Murani05} 
A. P. Murani, S. J. Levett, and J. W. Taylor, Phys. Rev. Lett {\bf 95},
256403 (2005).

\bibitem{Maddox06} 
B. R. Maddox, A. Lazicki, C. S. Yoo, V. Iota, M. Chen, A. K. McMahan,
M. Y. Hu, P. Chow, R. T.  Scalettar, and W. E. Pickett,
Phys. Rev. Lett. {\bf 96}, 215701 (2006).


\bibitem{Rueff06}
J.-P. Rueff, J.-P. Iti\'e, M. Taguchi, C. F. Hague, J.-M. Mariot,
R. Delaunay, J.-P. Kappler, and N.  Jaouen, Phys. Rev. Lett. {\bf 96},
237403 (2006)


\bibitem{Johansson74}
B. Johansson, Philos. Mag. {\bf 30}, 469 (1974).

\bibitem{Johansson75}
B. Johansson, Phys. Rev. B {\bf 11}, 2740 (1975).

\bibitem{KVC1}
J. W. Allen and R. M. Martin, Phys. Rev. Lett. {\bf 49}, 1106 (1982).

\bibitem{KVC2}
M. Lavagna, C. Lacroix, and M. Cyrot, Phys. Lett. {\bf 90A}, 210 (1982).

\bibitem{KVC3}
L. Z. Liu, J. W. Allen, O. Gunnarsson, N. E. Christensen, and
O. K. Andersen, Phys. Rev.~B {\bf 45}, 8934 (1992).

\bibitem{KVC4}
J. W. Allen and L. Z. Liu, Phys. Rev.~B {\bf 46}, 5047 (1992).


\bibitem{Eriksson90} 
O. Eriksson, M. S. S. Brooks, and B. Johansson, Phys. Rev. B {\bf
41}, R7311 (1990).

\bibitem{Sandalov95} 
I. S. Sandalov, O. Hjortstam, B. Johansson, and O. Eriksson,
Phys. Rev. B {\bf 51}, 13987 (1995).

\bibitem{Shick01} 
A. B. Shick, W. E. Pickett, and A. I. Liechtenstein,
J. Electron Spectrosc. {\bf 114}, 753 (2001).

\bibitem{SoderlindPr02} 
P. S\"oderlind, Phys. Rev. B {\bf 65}, 115105 (2002).

\bibitem{SzotekCe94} 
Z. Szotek, W. M. Temmerman, and H. Winter, Phys. Rev. Lett.
{\bf 72}, 1244 (1994).

\bibitem{SvaneCe94} 
A. Svane, Phys. Rev. Lett. {\bf 72}, 1248 (1994); Phys. Rev. B
{\bf 53}, 4275 (1996).

\bibitem{SvanePr97} 
A. Svane, J. Trygg, B. Johansson, and O. Eriksson Phys. Rev. B
{\bf 56}, 7143 (1997).

\bibitem{1N} 
O. Gunnarsson and K. Sch\"onhammer, Phys. Rev. Lett. {\bf 50}, 
604 (1983); Phys. Rev. B {\bf 28}, 4315 (1983); 
Phys. Rev. B {\bf 31}, 4815 (1985). 

\bibitem{LDADMFT1} 
K. Held, I. A. Nekrasov, G. Keller, V. Eyert, N. Blumer, A. K. McMahan,
R. T. Scalettar, T. Pruschke, V.  I. Anisimov, and D. Vollhardt, Phys. Status
Solidi B {\bf 243}, 2599 (2006)

\bibitem{LDADMFT2} 
G. Kotliar, S. Y. Savrasov, K. Haule, V. S. Oudovenko, O. Parcollet,
and C. A. Marianetti, Rev. Mod. Phys. {\bf 78}, 865 (2006).

\bibitem{DMFT1}
D. Vollhardt, in {\it Correlated Electron  Systems}, edited by
V. J. Emery (World Scientific, Singapore) 57 (1993); Th. Pruschke
M. Jarrell, and J. K. Freericks, Adv. Phys. {\bf 44}, 187 (1995).

\bibitem{DMFT2}
A. Georges, G. Kotliar, W. Krauth, and M. Rozenberg,
Rev. Mod. Phys. {\bf 68}, 13 (1996).


\bibitem{Zoelfl01}
M. B. Z\"olfl, I. A. Nekrasov, Th. Pruschke, V. I. Anisimov,
and J. Keller, Phys. Rev. Lett. {\bf 87}, 276403 (2001).

\bibitem{Held01} K. Held, A. K. McMahan, and R. T. Scalettar
Phys. Rev. Lett. {\bf 87}, 276404 (2001).

\bibitem{McMahan03} A. K. McMahan, K. Held, and R. T. Scalettar,
Phys. Rev. B {\bf 67}, 075108 (2003).

\bibitem{McMahan05}
A. K. McMahan, Phys. Rev. B {\bf 72}, 115125 (2005).

\bibitem{Haule05}  
K. Haule, V. Oudovenko, S. Y. Savrasov, and G. Kotliar,
Phys. Rev. Lett. {\bf 94}, 036401 (2005).

\bibitem{Amadon06} 
B. Amadon, S. Biermann, A. Georges, and F. Aryasetiawan, Phys
Rev. Lett. {\bf 96}, 066402 (2006)

\bibitem{Koga05} 
A. Koga, N. Kawakami, T.M. Rice, and M. Sigrist,
Phys. Rev. {\bf B} 72, 045128 (2005).

\bibitem{Ylvisaker09} 
E. R. Ylvisaker, J. Kune\v{s}, A. K. McMahan, and W. E. Pickett,
Phys. Rev. Lett. {\bf 102}, 246401 (2009).

\bibitem{Hirsch86}
J. E. Hirsch and R. M. Fye, Phys. Rev. Lett. {\bf 56}, 2521 (1986).

\bibitem{QMC1bnd}
See Ref. \onlinecite{DMFT2} and M. Jarrell, in {\it Numerical
Methods for Lattice Quantum Many-Body Problems}, editor D. Scalapino
(Addison Wesley, 1997) for one-band DMFT(QMC).

\bibitem{Jarrell96}  
M. Jarrell and J. E. Gubernatis, Phys. Reports {\bf 269}, 133 (1996).

\bibitem{young91} 
D. A. Young, {\it Phase Diagrams of the Elements} (University of
California Press, Berkeley, 1991).

\bibitem{nfvsV}
Fig.~2 in Ref.~\onlinecite{McMahan05} shows $n_f\!\gtrsim\!n$
for Ce ($n\!=\!1$), Pr (2), and Nd (3), with $n_f$ monotonically
increasing under compression except for Ce in the range $24\!\leq
\! V \! \leq \! 29\,$\AA$^3$/atom.

\bibitem{footnote}
For a single-band Hubbard model at fixed filling $n=1$, the weights
$w_{n-1}=w_{n+1}$
contain identical information concerning possible localization
transitions.
One can partially restore this symmetry even in the present
multi-band case, when the filling drifts away from integer values,
by subtracting off the excess density.  In particular,
the identities in Eq.~\ref{eqnstatw} yield
$w_{n+1} - [ n_l - n ]
= w_{n+1} - [ (n-1) w_{n-1} + n w_n +(n+1) w_{n+1} ] 
= w_{n+1} - [ -w_{n-1} + w_{n+1} ]
= w_{n-1} $.
Note, however, that the asymmetry between $f^1-f^0$ and $f^2-f^1$ mixing
described in Ref.~\onlinecite{KVC4}
suggests this attempt to make the two weights equivalent should
not be pushed too strongly.

\bibitem{Nd10}
Our LDA estimate of the Nd pressure at $10\,$\AA$^3$/atom is $365\,$GPa.

\bibitem{Lipp08} 
M. J. Lipp, D. Jackson, H. Cynn, C. Aracne, W. J. Evans, and
A. K. McMahan, Phys. Rev. Lett. {\bf 101}, 165703 (2008).

\bibitem{Jeong04} 
I.-K. Jeong, T. W. Darling, M. J. Graf, Th. Proffen, R. H. Heffner,
Yongjae Lee, T. Vogt, and J. D. Jorgensen Phys. Rev. Lett. 92, 105702
(2004).



\end{thebibliography}
\end{document}